\documentstyle[12pt]{article}

\begin{document}
\title{\textbf{Slow dynamos in Lorentz tori Anti-de Sitter spacetime embedded in Riemann 2D-space}} \maketitle
{\sl \textbf{L.C. Garcia de Andrade}\newline
Departamento de F\'{\i}sica
Te\'orica-IF\newline
Universidade do Estado do Rio de Janeiro\\[-3mm]
Rua S\~ao Francisco Xavier, 524\\[-3mm]
Rio de Janeiro, RJ, Brasil\\[-3mm]
\\[-3mm]
\vspace{0.01cm} \newline{\bf Abstract} \paragraph*{Earlier Chicone, Latushkin and Montgomery-Smith [Comm Math Phys (1997)] have shown that a fast dynamo in compact two-dimensional manifold can be supported as long as its Riemannian curvature be negative. Recently Klebanov and Maldacena [Phys Today (2008)] showed that a similar flat spacetime embedding of a 2D negative Riemannian hyperbolic embedding in 2+1-D space-time, is achieved by a coordinate transformation. This embedding is used here to obtain a flat spacetime embedding of a slow dynamo in Riemannian 2D compact manifold of negative constant curvature. In is shown that a slow dynamo appears in anti-de Sitter space (AdS) Lorentz tori. This is in agreement with Bassett et al [Phys Rev D (2001)] cosmic dynamo where suppression of resonance by universe expansion slow dynamo action in comparison to preheating phases. Other example of flat embeddings, which keeps some resamblance with AdS slow dynamo, is given by the embedding of Moebius strip [Shukurov, Stepanov, Sokoloff, PRE (2008)] in the laboratory. A simplest one is given by the embedding of twisted flux tubes in the 3D Euclidean space. Decaying of magnetic fields are shown analytically, for non-Moebius strip surfaces. Shukurov et al have also found slow dynamos numerically. Actually it is easy to solve the induction equation in Euclidean space and obtain the induced Moebius flow magnetic fields by Jacobian transformation. The same technique can be done to obtain the parabolic slow dynamo.}
\newpage
\section{Introduction}
 Earlier Chicone and \cite{1} called the attention to the fact that there are certain flat and negative curvature embeddings of Moebius strip into the $3D$ real Euclidean space. Actually they stablished very simple criteria to this flat embedding based on the torsion of the embedded space. Here not only this criteria may be used in solar flux tubes, as we shall show in section II but it has been also used to embedded a Moebius dynamo flow in a torus device called the liquid sodium Perm dynamo experiment \cite{2}. Here it is shown that for a non-Moebius thin surface with ellipsoidal cross-section the magnetic field embedded in the $\textbf{R}^{3}$ laboratory, decays in a non-dynamo profile. Unfortunatly in some of these beautiful mathematical models of physical systems one main difficult is the embedding problems of the 2D negative constant Riemann curvature surfaces in 3D Euclidean spaces, due to Hilbert theorem which states that the embedding of negative curvature spaces in $\textbf{R}^{3}$ would have at least one point of positive curvature and therefore the constancy of the negative curvature would be spoiled. A way out of this problem has been suggested recently \cite{3}. This idea is to embedded negative Riemannian curvature space in cosmology where no further embedding is required. This has been recently accomplished by Klebanov and Maldacena \cite{4} in the context of high energy extra dimensions cosmology, which goes as follows. They simple argued that by a suitable coordinate transformation one is able to reduce the $2+1$-D spacetime to a 2D negative constant curvature Riemannian hyperbolic space. The same reasoning is used by Carneiro \cite{5}, by considering the Friedmann-Goedel metric in the form of 2D negative curvature embedding into the spatial 3D section of the $3+1$-D space-time cosmology. It is shown that from Clarkson-Marklund \cite{6} GR MHD dynamo equation, a slow dynamo solution is obtained in the negative constant Riemannian compact manifold embedded into the 3D Euclidean space. In their equation an interesting aspect is that Ricci curvature only appears when diffusion is present. A detailed account of GR cosmological magnetic fields can be found in Widrow \cite{7}. The paper is organized as follows: Section II presents the embedding problem and the relation with the flat embedding of solar flux tubes which undergoes flat embedding in $\textbf{R}^{3}$. Section III deals with the embedding of magnetic fields in the Moebius strip flow which can be shown to be a dynamo flow, under certain constraints in the laboratory. Section IV, deals with the cosmological dynamos, and shows that the flat spacetime embedding leads natural to a slow dynamo in isotropic cosmology. AdS magnetic fields are shown to decay and slow dynamos can be obtained also from flat embedding in space of negative constant Riemann curvature. Section IV presents future prospects and conclusions.
\newpage
\section{Flat embedding of twisted magnetic flux tubes in $\textbf{R}^{3}$}
In this section, one shall be concerned with the problem of embedding of 2-surfaces in 3D Euclidean spaces. According to mathematicians this embedding can be done in two ways ,one is flat embedding and other is negative curved embedding. In this section one shall be concerned with the demonstration that some physical important spaces in dynamo theory can be flat embedding while in the next section one addresses the issue of negative curved embedding and this importance to cosmic dynamos. A simple criteria investigated by Chicone and \cite{8} for the flat embedding of a Moebius strip is to consider that the torsion integral
\begin{equation}
{\theta}(s)=constant+\int{{\tau}(s)ds}
\label{1}
\end{equation}
Let us consider the flat Euclidean 3D $\textbf{E}^{3}$ metric
\begin{equation}
ds^{2}=dx^{2}+dy^{2}+dz^{2}
\label{2}
\end{equation}
the embedding can be simply done by writing the retangular coordinates $(x,y,z)$ in terms of the coordinates $(u,v)$ of the embedded surfaces. For example, in the case of twisted magnetic flux tube dynamos in solar physics, the surface metric would be described by \cite{9}
\begin{equation}
{ds_{FT}}^{2}={r_{0}}^{2}d{{\theta}_{R}}^{2}+ds^{2}
\label{3}
\end{equation}
in the twisted flux tube surface coordinates $({\theta}_{R},s)$, where ${\theta}_{R}$ is the angular coordinate of the tube connected with the torsion integral by
\begin{equation}
{\theta}={\theta}_{R}-\int{{\tau}(s)ds}
\label{4}
\end{equation}
and the other coordinate s is the coordinate along the flux tube axis. Here the coordinate $r_{0}$ is the constant radial coordinate, which characterizes partially the surface. The transformation of the flux tube surface which characterizes the embedding is
\begin{equation}
x=r_{0}cos{{\theta}_{R}}
\label{5}
\end{equation}
\begin{equation}
y=r_{0}sin{{\theta}_{R}}
\label{6}
\end{equation}
and $z$ coincides locally with s-coordinate. Of course, the differentiation of coordinates $(x,y,z)$ above transforms Riemannian line element (\ref{3}) and the flat Riemann metric 3D in (\ref{2}). This characterizes also the flat embedding. A more general criteria for flat embedding of Moebius strips, which shall be used in the next section is given by
\begin{equation}
|\frac{d}{ds}{\theta}|=|{\tau}(s)|
\label{7}
\end{equation}
which can also obtained here for the flux tube twist angle ${\theta}(s)$. Though this first embedding example is quite simple, in the next section one shall be concerned with the embedding of much more complex surfaces, which include flat embeddings of Moebius and non-Moebius surfaces.
\section{Flat embedding of Moebius dynamo flows in $\textbf{R}^{3}$}
In this section it is shown that a simple method of embedding of Moebius strip dynamo flow in the laboratory Euclidean space. This dynamo action in Moebius flow has been obtained originally by Shukurov et al \cite{2}, from the numerical simulation of the self-induced equation for the Moebius strip background. They basically found a slow dynamo from the non-orthogonal coordinates $(u,v)$. Here one shall consider that a analytical solution can be found by embedding magnetic fields on the Moebius strip inducing a three-dimensional magnetic field in $\textbf{E}^{3}$. The relation between the magnetic field components $B_{u}$ and $B_{v}$ can then be mapped into the 3D magnetic field components $(B_{x},B_{y},B_{z})$, by the Jacobian transformation equation
\begin{equation}
B_{x}=({\partial}_{u}x)B_{u}+({\partial}_{v}x)B_{v}
\label{8}
\end{equation}
\begin{equation}
B_{y}=({\partial}_{u}y)B_{u}+({\partial}_{v}y)B_{v}
\label{9}
\end{equation}
\begin{equation}
B_{z}=({\partial}_{u}z)B_{u}+({\partial}_{v}z)B_{v}
\label{10}
\end{equation}
Here, the retangular coordinates ate given in terms of the $(u,v)$ coordinates by the embedding transformation
\begin{equation}
{x}(u,v)=(R+vcosnu)cos{u}
\label{11}
\end{equation}
\begin{equation}
{y}(u,v)=(R+vcosnu)sin{u}
\label{12}
\end{equation}
\begin{equation}
{z}(u,v)={v}sin(nu)
\label{13}
\end{equation}
The Riemann metric components $g_{ik}$ (i,k=1,2,3) of the Moebius flow is
\begin{equation}
g_{uu}=n^{2}v^{2}sin^{2}nu+(R+vcosnu)^{2}
\label{14}
\end{equation}
\begin{equation}
g_{uv}=-nv sin(nu)cos(nu)
\label{15}
\end{equation}
\begin{equation}
g_{vv}=cos^{2}(nu)
\label{16}
\end{equation}
Equations (\ref{8,9,10}) can be expressed as
\begin{equation}
B_{u}=\frac{-c}{(a sin{u}-b cosu)}B_{v}
\label{17}
\end{equation}
and
\begin{equation}
B_{z}=[\frac{-c nv cosnu}{(a sin{u}-b cosu)}+sinu]B_{v}
\label{18}
\end{equation}
where a,b and c are defined by
\begin{equation}
a=n v sin(nu)
\label{19}
\end{equation}
\begin{equation}
c= cosnu cosu
\label{20}
\end{equation}
\begin{equation}
b=(R+nv cosnu)\label{21}
\end{equation}
As a first physical implication of these expressions for the dynamo flow, one realizes that for $n=1$, for thin Moebius strips, $(R\approx{0})$ the dynamo ratio $\frac{B_{u}}{B_{v}}$ vanishes and no dynamo action exists, in the sense poloidal magnetic field $B_{v}$ are not converted into toroidal fields $B_{u}$. However note that when $R\ne{0}$ ellipsoidal cross-section Moebius strips are consider, the embedding leads to the following dynamo action
\begin{equation}
B_{u}=\frac{R sin{u}}{cos^{2}u}B_{v}
\label{22}
\end{equation}
Now let us consider a simple solution of the self-induction equation
\begin{equation}
\frac{{\partial}\textbf{A}}{{\partial}t}+{\eta}{\nabla}^{2}\textbf{A}=\textbf{U}\times{\nabla}
\textbf{A}\label{23}
\end{equation}
where $\textbf{A}$ is the magnetic vector potential whose magnetic field is given by
\begin{equation}
\textbf{B}={\nabla}\times\textbf{A}\label{24}
\end{equation}
Since the frame of reference in this case is Euclidean this equation reduces to the following component equation
\begin{equation}
\frac{{\partial}{A^{i}}}{{\partial}t}+{\eta}{\nabla}^{2}{A^{i}}=[U_{j}(A^{i,j}-A^{j,i})]
\label{25}
\end{equation}
where ${\eta}\approx{Rm^{-1}}$ is the diffusion related with the magnetic Reynolds number $Rm$. Note that the Laplacian term is given in covariant coordinates by
\begin{equation}
{\nabla}^{2}A^{i}={\eta}{A^{i;k}}_{;k}\label{26}
\end{equation}
where the Riemannian covariant derivative Laplacian ${A^{i;k}}_{;k}$ is given by
\begin{equation}
{A^{i;k}}_{;k}=[g^{jk}({A^{i}}_{,k}+{{\Gamma}^{i}}_{lk}A^{l})]_{,j}+g^{jl}{{\Gamma}^{i}}_{
kj}({A^{k}}_{,l}+{{\Gamma}^{k}}_{lm}A^{m})+g^{kl}{{\Gamma}^{j}}_{
kj}({A^{i}}_{,l}+{{\Gamma}^{i}}_{lm}A^{m})\label{27}
\end{equation}
one also considers here that the Coulomb gauge
\begin{equation}
{\nabla}.\textbf{A}=\frac{{\phi}}{\eta}=0\label{28}
\end{equation}
since the electric field potential ${\phi}$ is assumed to vanish. Expansion of the equation in terms of the Riemann-Christoffel symbols ${\Gamma}_{ijk}$ above, and assumption that $A^{z}$ vanishes, reduces the self-induction equation to
\begin{equation}
\frac{{\partial}{A^{x}}}{{\partial}t}=[U_{y}(A^{x,y}-A^{y,x})]
\label{29}
\end{equation}
and
\begin{equation}
\frac{{\partial}{A^{y}}}{{\partial}t}=[U_{x}(A^{y,x}-A^{x,y})]
\label{30}
\end{equation}
A combination of these two equations yields
\begin{equation}
\frac{{\partial}[{A^{y}}-\frac{{U}_{x}}{{U}_{y}}A^{x}]}{{\partial}t}=0
\label{31}
\end{equation}
Thus by finding a simple solution of these equations embedded in Euclidean 3D space allows us to obtain the hyperbolic dynamo magnetic field $(B_{u},B_{v})$. Note that a simple solution of this equation can be given by
\begin{equation}
B_{z}=[1+\frac{{{V}_{x}}^{2}}{{{V_{y}}}^{2}}]{\partial}_{x}{A^{y}}
\label{32}
\end{equation}
for a steady 2D flow $V_{x}(y)$ and $V_{y}(x)$. Thus by substitution of this expression into the Jacobian transformation above with $B_{x}$ zero yields $B_{u}$ and $B_{v}$ determined. In this case actually a decaying of the magnetic field is observed. This is actually in accordance with Cowling anti-dynamo theorem. These decays are very common also in cosmology. In next section one shall consider the curved embedding of dynamo cosmology.
\section{Curved and flat spacetime embedding of 2D slow dynamos}
Earlier Brandenburg et al \cite{10} have consider a two dimensional slice of expanding Friedmann universe with decaying magnetic fields from MHD turbulence. In this section it is shown that by considering a simple dynamo solution of Clarkson-Marklund equation in the Riemannian 2D compact manifold which is actually curved embedded into the $(3+1)-D$ cosmological model as given by Carneiro transformation
\begin{equation}
e^{x}= cosh{\epsilon}+cos{\phi}sinh{\epsilon}
\label{33}
\end{equation}
\begin{equation}
ye^{x}=sin{\phi}sinh{\epsilon}
\label{34}
\end{equation}
which is able to transform the anisotropic universe metric
\begin{equation}
ds^{2}=a^{2}({\eta}_{0})[d{{\eta}_{0}}^{2}-(dx^{2}+e^{x}dy^{2}+dz^{2})]\label{35}
\end{equation}
obtained from the Goedel metric in the radiation era, into the metric Riemannian line element of where ${\eta}_{0}$ is the conformal time. Zero index was placed not to confuse the reader with the diffusion constant above. Recently Klebanov and Maldacena \cite{3} have shown that a similar transformation given by
\begin{equation}
X= cos{\phi}sinh{\rho}
\label{36}
\end{equation}
\begin{equation}
Y=sin{\phi}sinh{\rho}
\label{37}
\end{equation}
\begin{equation}
Z=cosh{\rho}
\label{38}
\end{equation}
embedds the flat $(2+1)-D$ spacetime metric
\begin{equation}
ds^{2}=dX^{2}+dY^{2}-dZ^{2}
\label{39}
\end{equation}
into the intrinsic 2D hyperbolic Riemannian line element of negative constant curvature. In cosmology Anti-de Sitter space (AdS) of metric
\begin{equation}
ds^{2}=-e^{2y}dx^{2}+dy^{2}\label{40}
\end{equation}
can be embedded into the space
\begin{equation}
ds^{2}=-dX^{2}-dY^{2}+dZ^{2}\label{41}
\end{equation}
according to the coordinate transformations
\begin{equation}
X= cos{\phi}sinh{\rho}
\label{42}
\end{equation}
\begin{equation}
Y=sin{\phi}sinh{\rho}
\label{43}
\end{equation}
\begin{equation}
Z=cosh{\rho}
\label{44}
\end{equation}
Let us now compute the magnetic field from the AdS metric from the solenoidal equation as
\begin{equation}
B^{y}=B_{0}(t)e^{y}\label{45}
\end{equation}
Now by solving the Clarkson-Marklund MHD-GR dynamo equation
\begin{equation}
\dot{\textbf{B}}(1+\frac{5}{3}{\Theta}{\eta})-(1+\frac{2}{3}{\Theta})\frac{2}{3}
{\Theta}\textbf{B}=
{\nabla}{\times}(\textbf{V}{\times}\textbf{B})+{\eta}{\Delta}\textbf{B}+
{\eta}<\textbf{Ric},\textbf{B}>\label{46}
\end{equation}
Taking into account that the expansion ${\Theta}$ has to be computed in terms of the covariant derivative
\begin{equation}
{\Theta}={\nabla}_{A}V^{A}={\partial}_{A}V^{A}-{\Gamma}_{A}V^{A}\label{47}
\end{equation}
where ${\Gamma}_{A}$ is the trace of the Riemann-Christoffel symbols
\begin{equation}
{{\Gamma}^{A}}_{BC}=\frac{1}{2}g^{AD}(g_{DB,C}+g_{DC,B}-g_{BC,D})
\label{48}
\end{equation}
In highly conducting the solution is simply
\begin{equation}
\dot{\textbf{B}}-\frac{2}{3}{\Theta}\textbf{B}=0
\label{49}
\end{equation}
and the solution is
\begin{equation}
{\gamma}=-2V^{y}=-2e^{-y}\le{0}
\label{50}
\end{equation}
This means that the magnetic field decays. Let us now consider the same technique of obtaining the magnetic fields on the embedded curved space by the Jacobian transformation, to the constant negatively curved Riemann compact two-dimensional space representing a parabolic flow. This is flat spacetime embedded by the coordinate transformation $(\ref{42,43,44})$ which shows that the magnetic fields in 2D parabolic geometry and flat (2+1)-D spacetime can be given by
\begin{equation}
B_{X}=({\partial}_{\rho}X)B_{\rho}+({\partial}_{\phi}X)B_{\phi}
\label{51}
\end{equation}
\begin{equation}
B_{Y}=({\partial}_{\rho}Y)B_{\rho}+({\partial}_{\phi}Y)B_{\phi}
\label{52}
\end{equation}
\begin{equation}
B_{Z}=({\partial}_{\rho}Z)B_{u}+({\partial}_{\phi}Z)B_{v}
\label{53}
\end{equation}
Since Z here represents the time component of spacetime the $B_{Z}$ component vanishes and the problem reduces to the following equations with $B_{\rho}$ zero, 
\begin{equation}
B_{X}=-sinh{\rho}.sin{\phi}B_{\phi}
\label{54}
\end{equation}
\begin{equation}
B_{Y}=sinh{\rho}.cos{\phi}B_{\phi}
\label{55}
\end{equation}
This yields the constraint relation that helps us to infer about the geometry of the magnetic fields
\begin{equation}
{B_{\phi}}^{2}=\frac{({B_{X}}^{2}+{B_{Y}}^{2})}{sinh^{2}{\rho}}
\label{56}
\end{equation}
Nevertheless to have a better idea of the magnetic field in the flat spacetime geometry one needs first to obtain the expression for the magnetic field $B_{\phi}$. Actually this AdS magnetic field seems to be similar in geometry to the Moebius strip one, namely in ellipsoid format. Let us consider the ansatz
\begin{equation}
{B_{\phi}}=\frac{{\sqrt{2}}{B_{0}}}{sinh{\rho}}e^{iKZ-{\lambda}^{2}{\eta}t}\label{57}
\end{equation}
Substituting this expression into (\ref{56}) one obtains
\begin{equation}
{({B_{X}}^{2}+{B_{Y}}^{2})^{\frac{1}{2}}}={{\sqrt{2}}{B_{0}}}e^{-{\eta}{\lambda}^{2}Z}
\label{58}
\end{equation}
From this expression one notes that the magnetic field oscillates along the Z direction but decays in time and besides the amplitude of the radius of the magnetic field circle collapses as time goes to infinite, which may consider the universe expansion. In the 2D parabolic geometry embedded in the flat (2+1)-D spacetime there is a ring like magnetic field, given explicitly only in terms of the $({\rho},{\phi})$ coordinates by
\begin{equation}
{B_{\phi}}=\frac{{\sqrt{2}}{B_{0}}}{sinh{\rho}}e^{-{\lambda}^{2}{\eta}cosh{\rho}}\label{59}
\end{equation}
where here K represents a constant wave wave number. Previous expression shows us that the Euclidean components $B_{X}$ and $B_{Y}$ depends only on the time coordinate Z, having therefore a Minkowski spacetime conic geometry. Actually this is called a Lorentz tori \cite{11}. Simply boundary conditions imposed by this solution shows that as ${\rho}\rightarrow{0}$ the magnetic field vanishes while as ${\rho}\rightarrow{\infty}$ the magnetic field of the ring would increase without bound spatially, but this magnetic field grow would quickly suppressed or compensated by the exponentially decay in time. Since the parabolic geometry is assumed to be bound is highly improbable that dynamo action can be fast. This ansatz satisfy the self-induction equation for the case of force-free Beltrami magnetic fields
\begin{equation}
{\nabla}{\times}\textbf{B}={\lambda}\textbf{B}\label{60}
\end{equation}
and the divergence-free equation. Besides one has assumed that the velocity flow is steady and that the flow is quasi-axi-symmetric where the magnetic and velocity fields are approximately parallel. The self-induction equation has the format
\begin{equation}
\frac{{\partial}{B_{X}}}{{\partial}Z}=-{\eta}{\lambda}^{2}B_{X}\label{61}
\end{equation}
with a similar equation for the $B_{Y}$ component. Here one must notice again that the coordinate Z substitutes the time coordinate t of previous section. Note from the above expressions that the magnetic energy density also decays and dynamo action cannot be supported. Of course due to Chicone et al proof of the existence of the fast dynamo in compact manifolds of negative constant Riemann curvature, one may hope that this non-dynamo result may be obtained from the absence of non-advection term and not from the low dimension of the space. \newpage
\section{AdS (1+1)-D pseudo-Riemannian geometry}
To confirm the last prediction, let us now compute the case of AdS spacetime in the case of diffusion where the Ricci tensor terms shall also be present. The geometrical quantities of the (1+1)-D AdS spacetime are the Ricci tensor components and scalar
\begin{equation}
R_{11}=-e^{-2y}\label{62}
\end{equation}
\begin{equation}
R_{22}=1\label{63}
\end{equation}
\begin{equation}
R=2\label{64}
\end{equation}
The Riemann-Christoffel symbols are
\begin{equation}
{{\Gamma}^{1}}_{21}=1\label{65}
\end{equation}
\begin{equation}
{{\Gamma}^{1}}_{22}=1\label{66}
\end{equation}
while the Riemann curvature tensor is 
\begin{equation}
{R_{1212}}=e^{-2y}\label{67}
\end{equation}
Let us now consider the slow dynamo solution of the GR dynamo equation of Clarkson-Marklund \cite{3} as
\begin{equation}
{\gamma}={\partial}_{x}lnB^{y}=-\frac{{\eta}[1+{\lambda}^{2}+\frac{2}{3}e^{-2y}]}{1-5{\eta}V^{0}e^{-2y}}
\label{68}
\end{equation}
where ${\gamma}$ is the rate of the amplification of the magnetic field. Thus since the limit
\begin{equation}
lim_{{\eta}\rightarrow{0}}\textbf{Re}{\gamma}(\eta)=0
\label{69}
\end{equation}
one can say that the dynamo is slow. Thus even when curvature and diffusion effects are taken into account, the quasi-axi-symmetric hypothesis of weakning the advection term yields a slow dynamo in negative Riemann constant curvature compact manifold embedded AdS space. 

\section{Conclusions}
       The importance of investigating the Riemannian geometrical aspects of the dynamo action of AdS spacetime embedded in the constant negative curvature compact Riemannian manifolds, stems from the facts that stretching of the AdS and its respective squashed studied recently by Bengtsson and Sandin \cite{11}, in the context of AdS black hole, are important in dynamo theory, besides it seems according to these authors that AdS spacetime had been found in the center of Milky way and in this aspect, this may be important for the galactic dynamos investigation under AdS structure Einstein-Weyl models for example. On the other hand the investigation of magnetic dilaton \cite{12} scalar fields in AdS geometry provides us with a new laboratory for dynamo action investigations in the universe and mechanisms for primordial magnetic fields to be amplified enough to predict the galactic dynamo fields observed in the present universe. Slow dynamos in the universe seems also to be in agreement with the suppresion of the amplification of the magnetic fields shown by Bassett et al \cite{12}.
       \section{Acknowledgements}
 Several discussions with J L Thiffeault are highly appreciated. I also thank financial  supports from UERJ and CNPq. Parts of this work started when the author was on two leaves of absence in Princeton university and in astrophysical dynamos program solar and stellar cycles and dynamos at NORDITA, Stockholm. Financial supports from both institutions are greatful acknowledged.
 
  \end{document}